\pdfoutput=1 

\documentclass{vldb}
\usepackage{graphicx}
\usepackage{algorithm,algorithmic,amsmath,graphicx,times,latexsym}
\usepackage{bbm} 
\usepackage{dsfont} 
\usepackage{subfigure}
\usepackage{color}
\newcommand{\cut}[1]{}

\def\DOM{\text{DOM}}
\begin{document}


\title{Scalable Probabilistic Databases with \\ Factor Graphs and MCMC}


%
%
%
%

\numberofauthors{3} 

\author{
%
%
\alignauthor
Michael Wick \\
       \affaddr{University of Massachusetts}\\
       \affaddr{Computer Science}\\
       \affaddr{140 Governor's Drive}\\
       \affaddr{Amherst, MA}\\
       \email{mwick@cs.umass.edu}
\alignauthor
Andrew McCallum\\
       \affaddr{University of Massachusetts}\\
       \affaddr{Computer Science}\\
       \affaddr{140 Governor's Drive}\\
       \affaddr{Amherst, MA}\\
       \email{mccallum@cs.umass.edu}
\alignauthor
Gerome Miklau\\
       \affaddr{University of Massachusetts}\\
       \affaddr{Computer Science}\\
       \affaddr{140 Governor's Drive}\\
       \affaddr{Amherst, MA}\\
       \email{miklau@cs.umass.edu}
}

\cut{
\additionalauthors{Additional authors: John Smith (The Th{\o}rv\"{a}ld Group,
email: {\texttt{jsmith@affiliation.org}}) and Julius P.~Kumquat
(The Kumquat Consortium, email: {\texttt{jpkumquat@consortium.net}}).}
\date{30 July 1999}
}

\maketitle

\begin{abstract}
  Probabilistic databases play a crucial role in the management and
  understanding of uncertain data. However, incorporating
  probabilities into the semantics of incomplete databases has posed
  many challenges, forcing systems to sacrifice modeling power,
  scalability, or restrict the class of relational algebra formula
  under which they are closed. We propose an alternative approach
  where the underlying relational database always represents a single
  world, and an external factor graph encodes a distribution over
  possible worlds; Markov chain Monte Carlo (MCMC) inference is then
  used to recover this uncertainty to a desired level of fidelity. Our
  approach allows the efficient evaluation of arbitrary queries over
  probabilistic databases with arbitrary dependencies expressed by
  graphical models with structure that changes during inference. MCMC
  sampling provides efficiency by hypothesizing {\em modifications} to
  possible worlds rather than generating entire worlds from
  scratch. Queries are then run over the portions of the world that
  change, avoiding the onerous cost of running full queries over each
  sampled world. A significant innovation of this work is the
  connection between MCMC sampling and materialized view maintenance
  techniques: we find empirically that using view maintenance
  techniques is several orders of magnitude faster than naively
  querying each sampled world. We also demonstrate our system's
  ability to answer relational queries with aggregation, and
  demonstrate additional scalability through the use of
  parallelization.
\end{abstract}

\section{Introduction}
A growing number of applications output large quantities of uncertain
data. For example, sensor networks produce imprecise readings and
information extraction systems (IE) produce errorful relational
records. Despite their inevitable inaccuracies, these types of
automatic prediction systems are becoming increasingly important. This
is evident by the sheer number of repositories culled from the web by
IE systems: CiteSeer, REXA, DbLife, ArnetMiner, and Google
Scholar. Probabilistic databases (PDBs) are a natural framework for
storing this uncertain output, but unfortunately most current PDBs do
not achieve the difficult balance of expressivity and efficiency
necessary to support such a range of scalable real-world structured
prediction systems.

Indeed, there is an inherent tension between the expressiveness of a
representation system and the efficiency of query evaluation. Many
recent approaches to probabilistic databases can be characterized as
residing on either pole of this continuum.  For example, some systems
favor efficient query evaluation by restricting modeling power with
strict independence assumptions
\cite{dalvi04efficient,norbert97probabilistic,andritsos06clean}. Other
systems allow rich representations that render query evaluation
intractable for a large portion of their model family
\cite{koch08maybms,wang08bayesstore,sen07representing,sen08exploiting}. In
this paper we combine graphical models and MCMC sampling to provide a
powerful combination of expressive freedom and efficient query
evaluation over arbitrary relational queries.

Graphical models are a widely used framework for representing
uncertainty and performing statistical inference in a myriad of
applications, including those in computational biology
\cite{sontag08tightening}, natural language processing
\cite{lafferty01conditional}, computer vision \cite{weinman09scene},
information extraction \cite{poon07joint}, and data integration
\cite{wick08discriminative}. These models are becoming even more
accessible with the proliferation of many general purpose
probabilistic programming languages
\cite{richardson06markov,milch04blog,mccallum08factorie}. Factor
graphs are a particular type of representation for graphical models
that serve as an umbrella framework for both Bayesian networks and
Markov random fields, and are capable of representing any exponential
family probability distribution.

Unfortunately, graphical models have been largely overlooked as a
choice for representing uncertainty in probabilistic databases. In
rare cases, connections have been drawn between graphical models and
PDBs \cite{koch08maybms,dalvi08probabilistic}, and only recently have
they been used explicitly in either the representation
\cite{wang08bayesstore}, or in the mechanism for query evaluation
\cite{sen07representing,sen08exploiting}. However, these systems are
in practice severely limited by the \#$\mathcal{P}$-hard problem of
query evaluation and would not scale to the types of sophisticated
models and large data crucial to many real-world problems
\cite{poon07joint,culotta07first,wick08discriminative}.

We distinguish ourselves from these lines of work in several important
ways. First, we directly address the problem of intractable query
evaluation and propose an approximate any-time approach that scales
both to dense factor graphs and large amounts of data. Second, we
avoid the issue of closing factor graph semantics under relational
algebra operators giving us the ability to evaluate any relational
algebra query (including aggregation). Third, we evaluate our approach
on a difficult real-world information extraction problem on which
exact statistical inference is intractable (even in the graphical
model framework). We are able to achieve this with a query evaluation
technique based on MCMC sampling. This is in contrast to previous
sampling approaches, which use traditional {\sl generative} Monte
Carlo methods \cite{jihad05mystiq,jampani08mcdb}. The Monte Carlo
sampling method of MCDB \cite{jampani08mcdb} requires knowing the
normalization constant for each function; unfortunately, for general
factor graphs this problem is as difficult as computing marginals
(\#$\mathcal{P}$-hard). On the other hand MCMC samplers hypothesize
local changes to worlds, avoiding the need to know the
normalizer. Additionally, MCMC enables us to track tuples affected by
local changes and we exploit this information to efficiently
re-evaluate the queries|avoiding the need to re-run the full query
from scratch over each sampled world.

Indeed we demonstrate query evaluation on such a factor graph (where
computing the normalization constant is intractable) and show that our
MCMC sampler based on view maintenance techniques reduces running time
by several orders of magnitude over the simple approach of running the
full query over each hypothesized world. We also empirically
demonstrate our ability to scale these intractable models to large
datasets with tens of millions of tuples and show further scalability
through parallelization. Finally, we demonstrate our evaluator's
ability to handle aggregate queries.

After introducing related work, the rest of the paper is organized as
follows: first we describe our representation, introduce factor
graphs, and use information extraction as a running pedagogical
example and application of our approach (although it more generally
applies to other problems that can be modeled by factor graphs). We
then introduce query evaluation techniques, including the materialized
view maintenance approach advocated in this paper. Finally, we present
experimental results demonstrating scalability to both large data and
highly correlated PDBs.

\section{Related Work}

Because early theoretical work on incomplete data focuses largely on
algebras and representation systems (e.g.,
\cite{biskup81formal,im84incomplete}), it was only natural to extend
this line of thinking to probabilities
\cite{barbara92management,widom05trio,koch08maybms,green06models,dalvi04efficient}\cut{benjelloun06uldbs}. However,
this extension is quite difficult since the probabilities in query
results must include {\em expressions} derived from the confidence
values originally embedded in the database. Systems meeting these
theoretical conditions must overcome a set of challenges that are
often satisfied at the expense of modeling-power or understandability.

Although there is a vast body of work on probabilistic databases,
graphical models have largely been ignored until recently. The work of
Sen et al. \cite{sen07representing,sen08exploiting} casts query
evaluation as inference in a graphical model and BayesStore
\cite{wang08bayesstore} makes explicit use of Bayesian networks to
represent uncertainty in the database. While expressive, generative
Bayesian networks have difficulty representing the types of
dependencies handled automatically in discriminative models
\cite{lafferty01conditional}, motivating a database approach to linear
chain conditional random fields \cite{wang10declarative}. We, however,
present a more general representation based on factor graphs, an
umbrella framework for both Bayesian networks and conditional random
fields. Perhaps more importantly we directly address the problem of
scalable query evaluation in these representations|with an MCMC
sampler|whereas previous systems based on graphical models are
severely restricted by this bottleneck. Furthermore our approach can
easily evaluate any relational algebra query without the need to close
the graphical model under the semantics of each operator.

There has also been recent interest in using sampling methods to
estimate tuple marginals or rankings. For example, the MystiQ
\cite{jihad05mystiq} system uses samplers \cut{similar to Huffton
  races} to estimate top-$k$ rankings \cite{re07efficient}. Joshi and
Jermaine apply variance reduction techniques to obtain better sample
estimates \cite{joshi09sampling}. MCDB \cite{jampani08mcdb} employs a
generative sampling approach to hypothesize possible worlds. However,
these approaches are based on feed-forward Monte Carlo techniques and
therefore cannot take advantage of the Markovian nature of MCMC
methods. The MCDB system does use the concept of ``tuple bundles'' to
exploit overlap across possible worlds, but this approach is difficult
to implement because it requires custom query optimization code and
redefining operators over bundles of tuples (requiring over 20,000
lines of C++ code; in contrast our approach is able to treat the DBMS
as a blackbox and still exploit overlap between samples). Furthermore,
MCDB requires an additional pre-processing step to compute the
overlap. In MCMC sampling, the overlap is determined automatically as
a byproduct of the procedure. This allows our method to employ ideas
from DBMS view materialization technology to take advantage of the
overlap between possible worlds. To the best of our knowledge, we are
the first Markov-chain Monte Carlo sampler for estimating
probabilities in probabilistic databases \cite{wick09representing}. We
are also the first to combine graphical models and sampling techniques
into a single cohesive probabilistic database representation system.


\section{Representation}

In our approach, the underlying relational database always stores a
single possible world (a setting to all the random variables),
enabling us to run any relational algebra query. Database objects such
as fields, tuples, and attributes represent random variables, however,
the factor graph expresses complex statistical relationships between
them. As required, we can recover uncertainty to a desired level of
fidelity through Markov chain Monte Carlo (MCMC), which hypothesizes
changes to random variable values that represent samples of possible
worlds. As this underlying database changes, we execute efficient
queries on the modified portions of worlds and obtain an increasingly
accurate approximation of the probabilistic answers. Another advantage
of a graphical model approach is that it enables automatic learning
over the database|avoiding the need to tune weights by hand.

We begin by describing factor graphs and the well known {\em possible
  worlds} semantics, where the uncertain database is a set of possible
worlds $W$, and each $w\in W$ is a deterministic instance of the
uncertain DB. Following tradition, we endow $W$ with a probability
distribution $\pi : W \rightarrow [0,1]$ s.t. $\sum_{w\in W}\pi(w)=1$,
yielding a distribution over possible worlds.

\subsection{Factor Graphs}

In our approach $\pi$ is encoded by a factor graph, a highly
expressive representation that can encode any exponential family
probability distribution (including Bayesian and Markov
networks). Indeed their success in areas such as natural language
processing, protein design, information extraction, physics, and
machine vision attest to their general representational power. Factor
graphs can succinctly capture relationships between random variables
with complex dependencies, making them a natural choice for relational
data.

Mathematically, a factor graph (parametrized by $\theta$) is a
bipartite graph whose nodes consist of the pair
$\mathcal{G}_{\theta}=\left<V,\Psi\right>$ where $V=X \cup Y$ is the
set of random variables: $X$ is the set of observed variables, and $Y$
is the set hidden variables; $\Psi=\{\psi_k\}$ is the set of
factors. \cut{We use capital letters to denote sets of random
  variables (as in $X$), and subscripts to refer to particular
  variables as in $X_i\in X$. Furthermore, we use lowercase to denote
  a scalar value \cut{from the domain} (e.g., $x_i$), and use the
  notation $X_i=x_i$ to indicate that variable $X_i$ is taking on the
  value $x_i$. Finally, we use superscripts to denote sets (of arity
  represented by the superscipt): the notation $X^r=x^r$ means the set
  of variables $\{X_i,X_{i+1},\cdots,X_{i+r}\}$ take on the values
  $(X_i=x_i,X_{i+1}=x_{i+1},$ $\cdots,$ $X_{i+r}=x_{r+r})$ where it
  is implicitly assumed that $x_i$ is a value from $X_i$'s domain.} \\

\noindent {\bf Random Variables} \\
\noindent Intuitively, random variables represent the range of values
that an uncertain object in the database may acquire.  Each hidden
variable $Y_i \in Y$ is associated with a domain $\DOM(Y_i)$
representing the range of possible values for $Y_i$. For example, the
domain could be binary $\{$yes,no$\}$, enumerations $\{$tall, grande,
venti$\}$ or real-valued $\{r \in \Re|r \geq 4\}$.  Observed variables
are fixed to a particular value in the domain and can be considered a
constant. For simplicity, and without loss of generality, we will
assume that random variables are scalar-valued (vector and set valued
variables can be re-written as a combination of factors and
variables).

In our notation, capital letters with a subscript (e.g., $Y_i, X_i$)
represent a single random variable, and lowercase letters (e.g.,
$y_i$) represent a value from the corresponding variable's domain:
$y_i\in \DOM(Y_i)$. We use the notation $X_i=x_i$ to indicate that
variable $X_i$ is taking on the value $x_i$. Finally, we use
superscripts to denote sets (of arity represented by the superscipt):
the notation $X^r=x^r$ means the set of variables
$\{X_i,X_{i+1},\cdots,X_{i+r}\}$ take on the values
$(X_i=x_i,X_{i+1}=x_{i+1},$ $\cdots,$ $X_{i+r}=x_{r+1})$ where it is
implicitly assumed that $x_i$ is a value from $X_i$'s domain. Capital
letters without a subscript refer to the entire variable space ($Y$
is all hidden variables and $X$ is all observables).\\

\noindent {\bf Factors} \\
\noindent Factors model dependencies between the random variables. In
fact, multiple factors may govern the behavior of the same variable by
expressing preferences for certain assignments to that variable over
others. This flexible overlapping structure is powerful for modeling
real world relational data.

Formally, each factor $\psi: x^m \times y^n \rightarrow \Re_+$ maps
{\em assignments} to subsets of observed variables $x^m \subseteq
\DOM(X)$ and hidden variables $y^n \subseteq \DOM(Y)$ to a
non-negative real-valued scalar. Intuitively, factors measure the
compatibility of settings to a group of variables, providing a
measurement of the uncertainty that the particular assignment
contributes to the world. For an example, see
Figure~\ref{fig:pedagog}. \cut{For example, if $X_1$ is an observed
  variable with domain $\DOM(X_1)=\{red,blue,black\}$ and $Y_1$ is a
  hidden variable with domain $\DOM(Y_1)=\{yes,no\}$, then the factor
  would assign a real value for each combination of its arguments: the
  assignment $X_1=red, Y_1=yes$ can be compared to the assignment
  $X_1=red, Y_1=no$ through the factor $\psi(x \in \DOM(X_1),y \in
  \DOM(Y_1))$.}

Typically, factors are computed as a log-linear combination of a
sufficient statistic (or feature function) $\phi_k$ and corresponding
parameter $\theta_k$ as $\psi_k(x^{m},y^{n}) =
\exp\left(\phi_k(x^{m},y^{n})\cdot\theta_k\right)$. Where $\phi$ are
user-specified features for representing the underlying data and
$\theta$ are corresponding real-valued weights measuring each features
impact. There are a number of available methods from machine learning
and statistics for automatically determining these weights (avoiding
the need for manual tuning).

Given the above definitions, the factor graph $\mathcal{G}_\theta$
expresses a probability distribution (parametrized by $\theta$, and
conditioned on $X$) $\pi_{\mathcal{G}} : X \times Y \rightarrow [0,1]$
s.t. $\sum_{y \in \DOM(Y)}\pi_{\mathcal{G}}(y|x)=1$. More specifically,
if the graph decomposes into a set of factors $\Psi$ (where each
$\psi\in\Psi$ has a factor-specific arity of $s+t$) then the
probability distribution $\pi_{\mathcal{G}}$ is given as:

\begin{equation}
\label{eqn:conditional222}
\pi_{\mathcal{G}}(Y=y|X=x;\theta)=\frac{1}{Z_X}\prod_{\psi\in\Psi}\psi(y^{s},x^{t})
\end{equation}

\noindent where $Z_X=\sum_{y\in Y}\prod_{k=1}^{n}\psi_k(y^s,x^t)$ is
an input-dependent normalizing constant ensuring that the distribution
sums to $1$. Note two special cases: if $X$ is empty then $\mathcal{G}$
is a Markov random field, and when factors
are locally normalized $\mathcal{G}$ is a Bayesian network. \\

\subsection{Possible Worlds}
An uncertain database $\mathcal{D}$ is a set of relations $R=\{R_i\}$
each with schema $S_i^k$ (of arity $k$) containing attributes
$R_i.a_1,\cdots,R_i.a_k$. Each attribute is equipped with a finite
domain $\DOM(R_i.a_1)$ (a field is certain if its value is known,
otherwise it is uncertain). A deterministic tuple $t$ for relation
$R_i$ is a realization of a value for each attribute
$t=\left<v_1,\cdots,v_k\right>$ for constants $v_1\in \DOM(a_1)\cdots
v_k\in \DOM(a_k)$. Let $T$ be the set of all such tuples for all such
relations in the database. Then the set of {\em all} (unrestricted)
worlds realizable by this uncertain database is $W_{\mathcal{D}} =
\{w~|~w\subseteq T\}$. 

Let each field in the database be a random variable whose domain is
the same as the field's attribute's domain. A deterministic field is
an observed variable $X$ and an uncertain field is a hidden variable
$Y$. Because each field is interpreted as a random variable with a
domain equivalent to its attribute's, the hypothesis space of the
random variables ($X$ and $Y$) contain the set of possible
worlds. Deterministic factors can model constraints over arbitrary
sets of variables by outputting $1$ if the constraint is satisfied,
and $0$ if it is violated (rendering such a world world
impossible). We then formally define $W$ to be all possible worlds
with respect to the factor graph's probability distribution $\pi$:

\begin{equation}
\label{eqn:possible-worlds}
W=\{w \in W_{\mathcal{D}} ~|~ \pi_{\mathcal{G}}(w)>0\}
\end{equation}

\subsection{Example}
\label{sec:example}
\begin{figure*}
\begin{center}
\includegraphics[width=1.0 \textwidth]{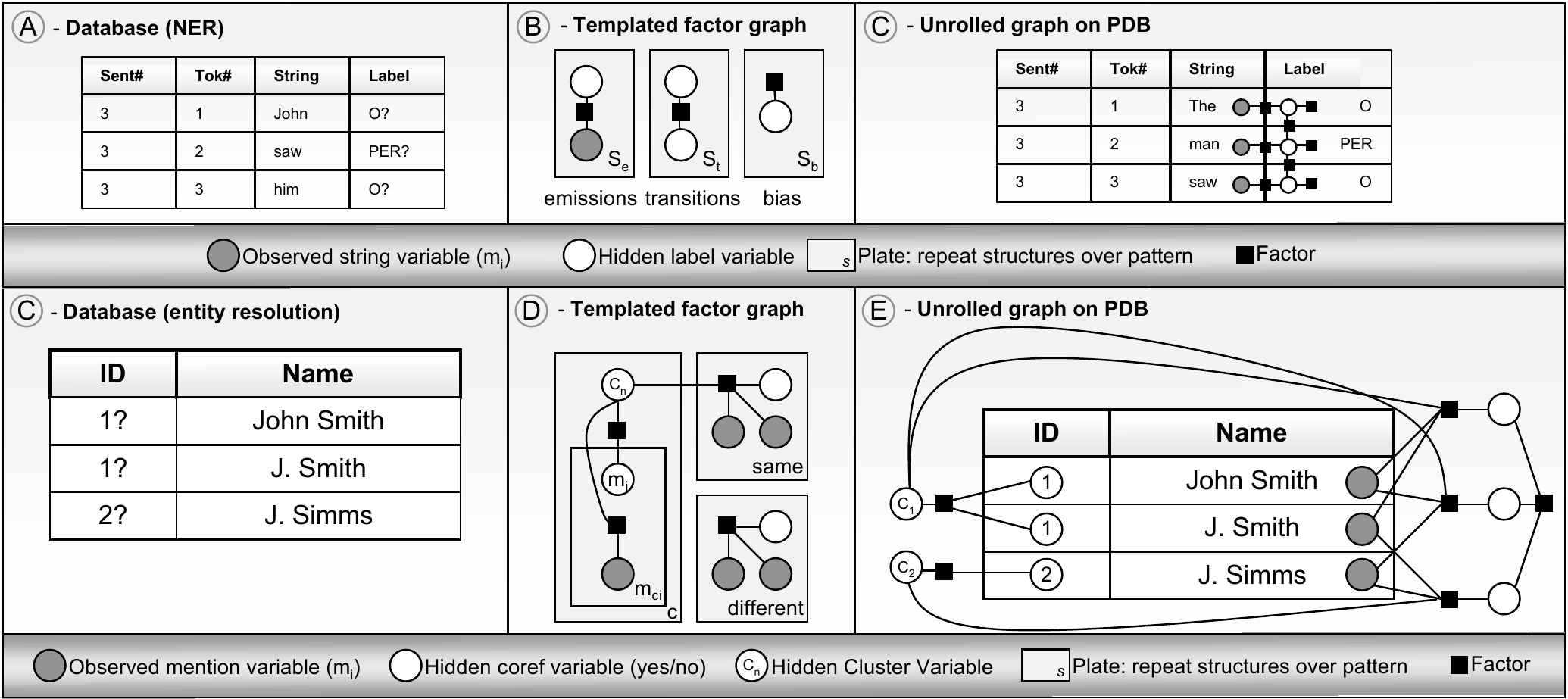}
\caption{Two information extraction problems in our framework. The top
  row is the problem of NER and the bottom row is problem of entity
  resolution. The first column (Panes A and C) shows a deterministic
  possible world for the respective problems. The second column (Panes
  B and D) shows the template specification for a factor graph for
  modeling these problems. The third column (Panes C and E) shows the
  templated graph unrolled on the possible world.}
\label{fig:pedagog}
\end{center}
\end{figure*}

We show two information extraction problems in
Figure~\ref{fig:pedagog} as represented in our approach. The top three
panes show named entity recognition (NER), and the bottom three panes
show entity resolution (disambiguation). NER is the problem of
identifying mentions of real-world entities in a text document; e.g.,
we might identify that ``Clinton'' is a {\em person} entity and
``IBM'' is an {\em organization} entity. The problem is usually cast
as sequence labeling, where each input sentence is divided into a
token sequence, and each word (token) in the sequence is treated as an
observed variable with a corresponding hidden variable representing
the label (entity type) that we are trying to predict. To model this
problem with a factor graph, we use factor templates to express
relationships between different types of random variables; in our NER
example, we express three such relationships (Pane B). The first is a
relationship between observed strings and hidden labels at each
position in the sequence (called the emission dependency: e.g., this
models that the string ``Clinton'' is highly correlated with the label
``person''). The second is a relationship between labels that neighbor
in the sequence (known as transition or 1st order Markov dependency:
for example, it is likely that a person label will follow a person
label because people have first and last names), the final dependency
is over each label, modeling the fact that some labels are more
frequent than others. Given the template specifications, the graph can
be unrolled onto a database. Pane C shows the random variables and
factors instantiated over the possible world initially shown in Pane
A. The probability of this world is simply a product of all the
factors (black boxes) illustrated in Pane C.

The bottom row of Figure~\ref{fig:pedagog} shows the problem of entity
resolution. Once mentions of named entities have been identified,
entity resolution clusters them into real-world entities. The database
in pane C shows a single possible world, the templated factor graph in
Pane D models relationships between the mentions, allowing
dependencies over entire clusters of mentions, dependencies between
mentions in the same cluster (modeling that mentions in clusters
should be cohesive), and dependencies between variables in different
clusters (modeling that mentions in separate clusters should be
distant). Finally, Pane E shows the graph unrolled on the database;
once again, the score of this possible world is proportional to the
product of factors in the unrolled graph. These examples simply serve
as an illustration, in practice we will exploit the benefits of MCMC
inference to {\em avoid instantiating the factor graphs over the entire
database}.

\subsection{Metropolis-Hastings}

Metropolis-Hastings (MH)
\cite{metropolis53equations,hastings1970monte} is an extremely general
MCMC framework used for estimating intractable probability
distributions over large state spaces. One advantage of MCMC is that
it can produce samples from the probability distribution $\pi$ without
knowledge of the normalization constant $Z_X$ (which is
\#$\mathcal{P}$-hard to compute). We will see in this section that
Metropolis-Hastings has many advantages, allowing us to avoid the need
to instantiate the graphical model over the entire database. The basic
idea is that MCMC can hypothesize changes to the single underlying
possible worlds by proposing modifications to previous worlds. We
describe MH more generally below.

MH requires two components, a target distribution that we wish to
sample (in our case $\pi(w)$) and a proposal distribution $q(\cdot|w)$
which conditioned on a state $w$ probabilistically produces a new
world $w'$ with probability $q(w'|w)$. The idea is that $q(\cdot|w)$
is a distribution from which we can easily sample (in practice such
distributions are easy to construct, and even allow us to inject
domain-specific knowledge).

The algorithm is initialized to a possible world $w_0$ (for example,
randomly). Next samples are drawn from the proposal distribution
$w' \sim q(\cdot|w)$, and each sample can either be accepted or
rejected according to a Bernoulli distribution given by parameter
$\alpha$:

\begin{equation}
\label{eqn:mh_accept}
\alpha(w',w) = \min\left(1, \frac{\pi(w')q(w|w')}{\pi(w)q(w'|w)}\right)
\end{equation}

\noindent The acceptance probability is determined by the product of
two ratios: the model probability ratio $\pi(w')/\pi(w)$ and the
proposal distribution ratio $q(w|w')/q(w'|w)$. Intuitively, the model
ratio captures the relative likelihood of the two worlds, and the
proposal ratio eliminates the bias introduced by the proposal
distribution. Given the requirement that the proposal distribution can
transition between any two worlds with positive probability in a
finite number of steps, the Metropolis-Hastings algorithm is
guaranteed to converge to the true distribution encoded by our factor
graph. Note that the normalization constant $Z$ appears in both the
numerator and denominator and cancels from the computation of
$\alpha$. Further, notice that only factors whose argument variables
are changed by $q$ need to be computed, and therefore only a small
portion of the graph needs to be unrolled on the database to evaluate
each proposal (for the two information extraction problems presented
in the previous section, a proposal that modifies only a constant
number of variables requires evaluating only a constant number of
factors). We show pseudo-code for performing a random-walk with MH in
the appendix: Algorithm~\ref{alg:mh}, and demonstrate how factors
cancel. We remark that another important advantage of MH is that it
avoids the need to explicitly enforce deterministic constraints
because the proposer $q$ is designed to transition within the space of
possible worlds only (in this sense $q$ is constraint-preserving). An
example of a constraint preserving proposal distribution is the
split-merge proposer for entity resolution, where clusters of mentions
are randomly split or merged (it is easy to check that these two
operations preserve the transitivity constraint: avoiding the need to
include the expensive cubic number of deterministic transitivity
factors).

\section{Query Evaluation}
\label{sec:query-evaluation}

The main query evaluation problem we are concerned with is to return
the set of tuples in the answer of a query $Q$ over the uncertain
database $\left<W,\pi\right>$, along with their corresponding
probabilities (of being in that answer). We say that a tuple $t$ is in
the answer of a query $Q$ if and only if $\exists w\in
W~\text{s.t.}~t\in Q(w)$. Then, the probability of this event is:

\begin{equation}
\label{eqn:poftexact}
Pr[t \in Q(W)] = \sum_{w \in W}{\mathds{1}_{t \in Q(w)}\pi(w)}
\end{equation}

\noindent We can see that if a tuple occurs in the answer for all
possible worlds, it is deterministic because
Equation~\ref{eqn:poftexact} sums to one. Similarly, a tuple occurring
in none of the deterministic answer sets has zero probability and
would be omitted from the answer. 

Unfortunately, Equation~\ref{eqn:poftexact} cannot be computed
tractably because it requires summing over the set of possible
worlds. Alternatively we can write the marginal probabilities as the
infinite-sample limit over a set of samples $S$ drawn from $\pi(\cdot|X)$:

\begin{equation}
\label{eqn:poft-sample}
Pr[t \in Q(W)] = \lim_{n\rightarrow\infty}\frac{1}{n}\sum_{i}^n\mathds{1}_{t \in Q(w_i\sim \pi(\cdot))}
\end{equation}

\noindent and estimate $Pr[t\in Q(W_{\mathcal{G}})]$ by using a finite
$n$. Given equation~\ref{eqn:poft-sample}, one approach is to draw
independent samples $w\sim_{\text{iid}}\pi$, requiring a generative
process that must completely instantiate each possible world (for
example, as done in MCDB \cite{jampani08mcdb}). However, generating a
possible world may be expensive in practice, motivating our approach
of using Markov-chain Monte Carlo to generate samples by equivalently
hypothesizing {\em modifications} to possible worlds.

There are two primary advantages of using a sampling approach for
estimating marginals. The first is that as $n$ goes to infinity, the
approximation becomes correct, allowing a trade-off between time and
fidelity: intuitively some applications are time sensitive and require
only course estimates of query marginals, while in others high
fidelity is extremely important. The second important property of
sampling methods is that they are query agnostic. That is, we need not
concern ourselves with closing the factor-graph representation over
every hypothetical query operator. For example, sampling methods
trivially handle aggregate extensions to relational algebra because
sampling from a graph returned as a query answer would be equivalent
to sampling from the original graph.

Up to this point, we have formally described our representation for
the possible worlds, the probability distribution, and have posed a
query evaluation problem of interest. We now focus our attention to
solving this query evaluation problem in our framework.  We first
overview background material, then describe a basic sampling
method. Finally, at the end of this section, we describe the main
algorithm of this article: Metropolis Hastings sampling with
materialized view maintenance.

\subsection{Basic MH Query Evaluation}
We now precisely define how to use Metropolis-Hastings to obtain
marginal probabilities for tuples in query answers. In particular, we
use Algorithm~\ref{alg:mh} to hypothesize a series of modifications to
worlds. Queries are then executed over hypothesized worlds, and the
marginal probabilities are computed according to
Equation~\ref{eqn:poft-sample} . We should note that consecutive
samples in MH are highly dependent; in situations such as ours, where
collecting counts is expensive (requires executing the query), it is
prudent to increase independence by collecting tuple counts only every
$k$ samples (a technique known as thinning). Choosing $k$ is an open
and interesting domain-specific problem. We present our basic MCMC
sampling method in Algorithm~\ref{alg:basic-sampling} (Appendix).

Another interesting scientific question is how to inject query
specific knowledge directly into the proposal distribution. For
example, a query might target an isolated subset of the database, then
the proposal distribution only has to sample this subset; this can be
(1) provided by an expert with domain-specific knowledge, (2)
generated by analyzing the structure of the graph and query, or even
(3) learned automatically through exploration. However, thoroughly
exploring this idea is beyond the scope of this paper.

Finally, there is an interesting balance between the traditional
ergodic theorems of MCMC and DBMS-sensitive cost issues arising from
disk-locality, caching, and indexing etc. For example, the ergodic
theorems imply that every MCMC sample be used to compute an
estimate. However, faced with the fact that each sample is non-trivial
to compute (requires executing a query), we must balance the
dependency of the samples with the expected costs of the
queries. Adaptively adjusting $k$ to respond to these various issues
is one type of optimization that may be applied to this problem.

\begin{algorithm}[h]
  \small
  \caption{Query Evaluation with Maintenance Techniques}\label{alg:efficient_view_refresh}
  \begin{algorithmic} [1]
    \STATE {\bf Input:} \\ initial world  $w_0$, \\ number of samples per query: $k$
    \STATE{{\bf Initialization:}
      \\ //run full query to get initial results\\$s \leftarrow$ Q$(w_0)$
      \\ //initial counts for marginals \\${\bf m}\leftarrow m_i=
    \begin{cases}
      1 & \text{if $m_i \in s$} \\
      0 & \text{o.w.} \\ 
    \end{cases}$
    \\ //initial normalizing constant for marginals
    \\ $z\leftarrow 1$}
    \\ $w\gets w_0$
    \FOR{$i=1,\hdots, $ number of steps}
       \STATE{$(w',\Delta^-,\Delta^+)\leftarrow $ MetropolisHastings($w$,$k$)}
       \STATE{$s \leftarrow s - Q'(w,\Delta^-) \cup Q'(w,\Delta^+)$}
       \STATE{
       ${\bf m}\leftarrow m_i+
       \begin{cases}
         1 & \text{if $m_i \in s$} \\
         0 & \text{o.w.} \\ 
       \end{cases}$}
       \STATE{$z\leftarrow z + 1$}
    \ENDFOR
    \STATE{{\bf return} $\frac{1}{z} {\bf m}$}
  \end{algorithmic}
\end{algorithm}

\subsection{MH Sampling with View Maintenance}
\begin{figure}[h]
\begin{center}
\includegraphics[width=0.28 \textwidth]{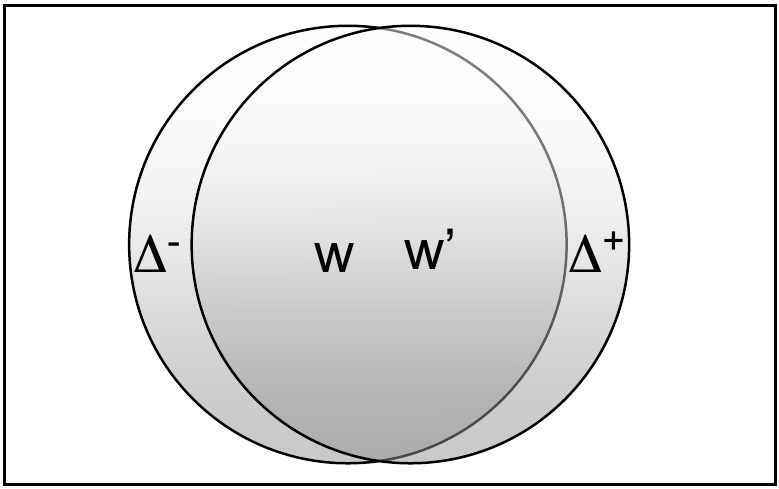}
\caption{$w$ is the original world, $w'$ is the new world after $k$
  MCMC steps and $\Delta^-\subseteq w$ is the set of tuples that were removed from $w$ and $\Delta^+\subseteq w'$ is the set added to $w'$.}
\label{fig:exploit_venn}
\end{center}
\end{figure}

Often, executing queries over a relational database is an expensive
resource consuming task. One way of obtaining tuple counts is to run
the query over each sampled world; however MCMC enables us to do much
better. Recall that consecutive samples in MCMC are actually
dependent; in fact, as illustrated in Figure~\ref{fig:exploit_venn}, a
world $w'$ is the result of a small modification to the original world
$w$.

We use this figure to directly motivate the relevance of materialized
view maintenance \cite{blakeley86efficiently}. Rather than run the
original (expensive) query over each consecutive sample, the query is
run only once on the initial world, then for each subsequent sample, a
modified query is run over the difference $\Delta$ and previous world
$w$. That is, we can exploit the semantics of set operations to obtain
an equivalent expression for the same answer set. Following the work
of Blakeley et al. \cite{blakeley86efficiently}, we recursively express the answer set as:

\begin{equation}
\label{eqn:recursive_efficient}
Q(w')=Q(w)-Q'(w,\Delta^-) \cup Q'(w,\Delta^+)
\end{equation}

\noindent where $Q'(w,\Delta^{\pm})$ is inexpensive because
$|\Delta^{\pm}|\ll|w|$ and $Q(w)$ is inexpensive because it can be
recursively expressed as repeated applications of
Equation~\ref{eqn:recursive_efficient} (bottoming out at the base case
of the initial world which is the only world that must be exhaustively
queried).

We discuss briefly some view materialization techniques. First,
observe that a selection $\sigma(w')$ can be re-written:

$$\sigma(w')\equiv \sigma(w) - \sigma(\Delta^-) \cup \sigma(\Delta^+)$$

\noindent and Cartesian products can similarly be re-written as:

\begin{align*}
w'.R_1\times w'.R_2 & \equiv w.R_1\times w.R_2 \\
& - w.R_1 \times \Delta^-.R_2 \\
& \cup w.R_1\times \Delta^+.R_2
\end{align*}

\noindent where $\Delta^-$ is the original setting of the tuples,
$\Delta^+$ is the new setting, and the notation $w.R_1$ is read as:
relation $R_1$ from world $w$. Traditional results from relational
algebra allow joins to be rewritten as a Cartesian product and a
selection. Further, it is not difficult to conceive how additional
relational operators such as various aggregators can be re-written in
terms of the sets $\Delta^-$ and $\Delta^+$.

In both the selection and join, the asymptotic savings can be as high
as a full degree of a polynomial (for example if $\Delta$ is constant
in size (as is often the case) and we lack indices over the fields
involved in the predicates). The high-level code for our
implementation based on view materialization techniques is exhibited
in Algorithm~\ref{alg:efficient_view_refresh}. In practice, the
implementation also requires the use of auxiliary tables for storing
the sets $\Delta^{-}$ and $\Delta^{+}$, which are necessary for
running the modified query $Q'$ from
Equation~\ref{eqn:recursive_efficient}. These tables must be updated during the course of Metropolis-Hastings, and additional cleaning and
refreshing of the tables and multi-set maps are required in between
deterministic query executions. \\

\noindent{\bf Remark:} please note that in the presence of projections
(as seen in all of our evaluation queries), that the set-difference
and set-union operators (from Equation~\ref{eqn:recursive_efficient})
actually requires multiset semantics, because counters need to be
maintained \cite{blakeley86efficiently}. We apply the necessary
modifications to Algorithm~\ref{alg:efficient_view_refresh} providing
additional book keeping to track the number of occurrences of each
tuple in the set $s$, so that the operators can be properly applied.

In practice, we handle projections in
Algorithm~\ref{alg:efficient_view_refresh} by maintaining the
multi-set maps from tuples to counts. In line 5, set difference and
set union are replaced with addition and subtraction operators to
maintain map counts, and in line 6, the condition is changed to:
$\text{count}(m_i)>0$.

\section{Experiments}
\label{sec:experiments}
In this section we demonstrate the viability of our approach by
representing the uncertain output of a real-world information
extraction problem: named entity recognition (NER). However, we go
beyond the simple linear-chain model and use a more sophisticated skip
chain conditional random field \cite{sutton04skip} to represent the
uncertainty in both the database and NER. Skip chains and similar
complex models achieve state-of-the-art performance in many IE tasks;
however no current PDB is capable of modeling them because exact
marginal inference is intractable. However, we demonstrate that our
MCMC approach effortlessly recovers the desired probabilities for
query evaluation.

We implement a prototype system in Scala \cite{odersky04scala}, a
functional extension to Java. The system is built in coordination with
our graphical model library for imperatively defined factor graphs,
Factorie, \cite{mccallum08factorie}, along with Apache Derby database
drivers interfaced through Java's JDBC API. We implement functionality
for (1) retrieving tuples from disk and then instantiating the
corresponding random variables in memory, and (2) propagating changes
to random variables back to the tuples on disk. Statistical inference
(MCMC) is performed on variables in main memory while query execution
is performed on disk by the DBMS. Additionally, we implement
infrastructure for both the naive and materialized-view maintenance
query evaluators. As random variables are modified in main memory,
their initial and new values are tracked and written to auxiliary
tables representing the ``added'' and ``deleted'' tuples required for
applying the efficient modified queries.

\subsection{Application: named entity recognition}
\label{sec:exp-ner}
We evaluate our probabilistic database on the real-world task of named
entity recognition. In particular we obtain ten-million tokens from
1788 New York Times articles from the year 2004. Recall that the
problem of named entity recognition is to label each token in the text
document with an entity type. We label the corpus using CoNLL
entities: ``PER'' (person entity such as Bill), ``ORG'' (organization
such as IBM), ``LOC'' (location such as New York City), ``MISC''
(miscellaneous entity|none of the above), and ``O'' (not a named
entity). We use BIO notation (see the appendix) to encode named
entities more than one token in length making the total number of
labels nine. We store the output of the ten million NYT tokens in a
database relation called TOKEN, with attributes ({\em
  \underline{TOK\_ID}, DOC\_ID, STRING, LABEL, TRUTH}) where {\em
  TOK\_ID} is underlined to indicate that it is the primary key, {\em
  DOC\_ID} is the document for which a token belongs, {\em STRING}
represents the text of a token, {\em LABEL} is unknown for all tuples
and is initialized to ``O'' , and {\em TRUTH} is a ``ground truth''
that we can use to train our model\footnote{to estimate ground truth
  we used the Stanford NER system (nlp.stanford.edu/ner/index.shtml)}.

\begin{figure}[h]
\begin{center}
\includegraphics[width=0.49 \textwidth]{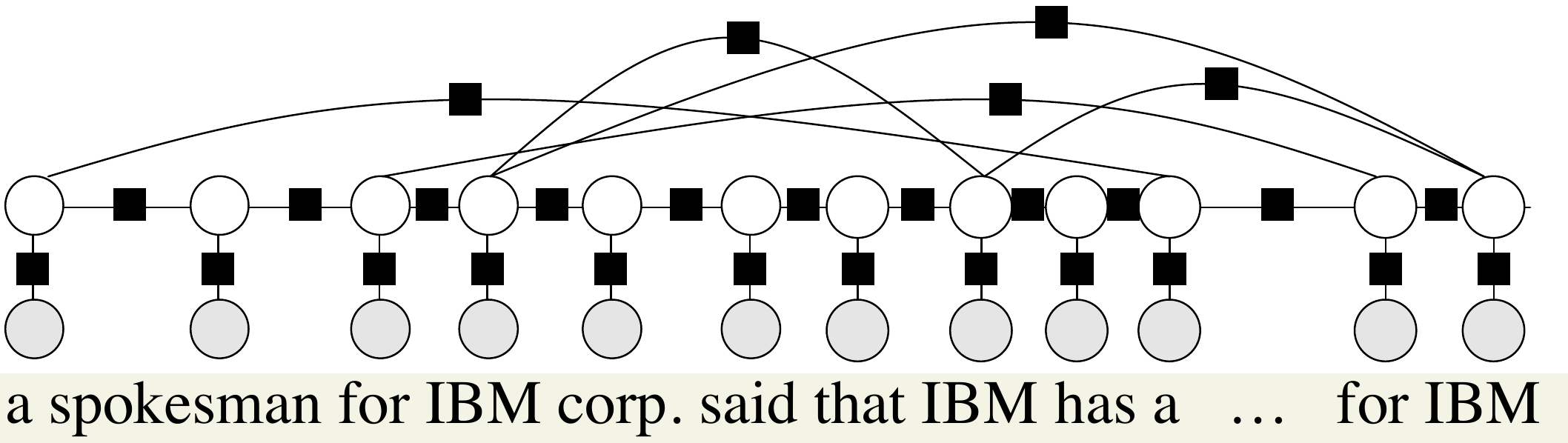}
\caption{A skip chain conditional random field that includes ``skip'' edges, or factors between tokens with the same string. Bias factors over labels are omitted for clarity.}
\label{fig:skip-chain}
\end{center}
\end{figure}

Next, we define the relational factor graph
(Figure~\ref{fig:skip-chain}) over the TOKEN relation to create our
probabilistic database. In particular, we first include the three
factor templates described in Section~\ref{sec:example}: (1) factors
between observed strings and corresponding labels, (2) transition
factors between consecutive labels, and (3) bias factors over
labels. Up to this point we have defined the traditional linear chain
model for NER (see \cite{lafferty01conditional}). However, skip-chain
models achieve much better results \cite{sutton04skip}, so we include
skip-edges or factors between labels whose strings are
identical. Intuitively, this factor captures the dependency that if
two tokens have the same string, then they have an increased
likelihood of having the same label. To see why inference in this
graph is intractable, note that the resulting factor graph
(Figure~\ref{fig:skip-chain}) is not tree-structured.

Now that we have defined our database and factor graph, we now define
our proposal distribution for query evaluation. Given a set of hidden
label variables $L$, our proposal distribution $q$ works as follows:
first a label variable is selected uniformly at random from $L$, then
the label for $L$ is randomly changed to one of the nine CoNLL labels
\{{\em B-PER, I-PER, B-ORG, I-ORG, B-MISC, I-MISC, B-LOC, I-LOSC,
  O}\}. This processes is repeated for 2000 proposals before $L$ is
changed by loading a new batch of variables from the database: up to
five documents worth of variables may be selected (documents are
selected uniformly at random from the database).

\subsection{Methodology}

We use the model and proposal distribution described in the previous
section in all experiments; we train the model using one-million steps
of SampleRank \cite{wick09samplerank}, a training method based on
MH. The method is extremely quick, learning all parameters in a matter
of minutes.  The query evaluation problems we investigate are all
instances of the general evaluation problem described in
Section~\ref{sec:query-evaluation}: the goal is to return each tuple
along with its probability of being in the answer set. We evaluate the
accuracy of our samplers by measuring the squared-error loss to the
ground truth query answer (that is, the usual element-wise squared
loss). Sometimes we report the normalized squared loss, which simply
scales the loss so that the maximum data point has a loss of $1$ (this
allows us to compare multiple queries on the same graph). Unless
otherwise stated, we estimate the ground-truth in each problem by
running our sampler for one-hundred-million proposals and collect a
sample every ten-thousand proposals. In all experiments we evaluate
the query every ten-thousand proposals (that is $k=10,000$ in
Algorithm~\ref{alg:basic-sampling}).

\subsection{Scalability}
In this section we demonstrate that we are able to scale query
evaluation to a large number of tuples even though exact inference in
the underlying graphical model (skip-chain CRF) is intractable and
approximate methods such as loopy belief propagation fail to converge
for these types of graphs \cite{sutton04skip}. We additionally compare
the materialized MCMC sampler with the basic (naive) MCMC sampler,
demonstrating dramatic efficiency gains. We use the following simple,
but non-selective query that scales linearly with the number of
tuples (note that the DBMS lacks an index over the STRING field): \\

\noindent {\bf Query 1} \\
{\small
\indent SELECT STRING\\
\indent FROM TOKEN \\
\indent WHERE LABEL='B-PER' \\
}

\begin{figure}[h]
\centering
\subfigure[Scalability over several orders of magnitude (Query~1); x axis is millions of tuples in log scale and y axis is time taken to half squared error.] 
{
    \label{fig:scalability-query}
    \includegraphics[width=0.44 \textwidth]{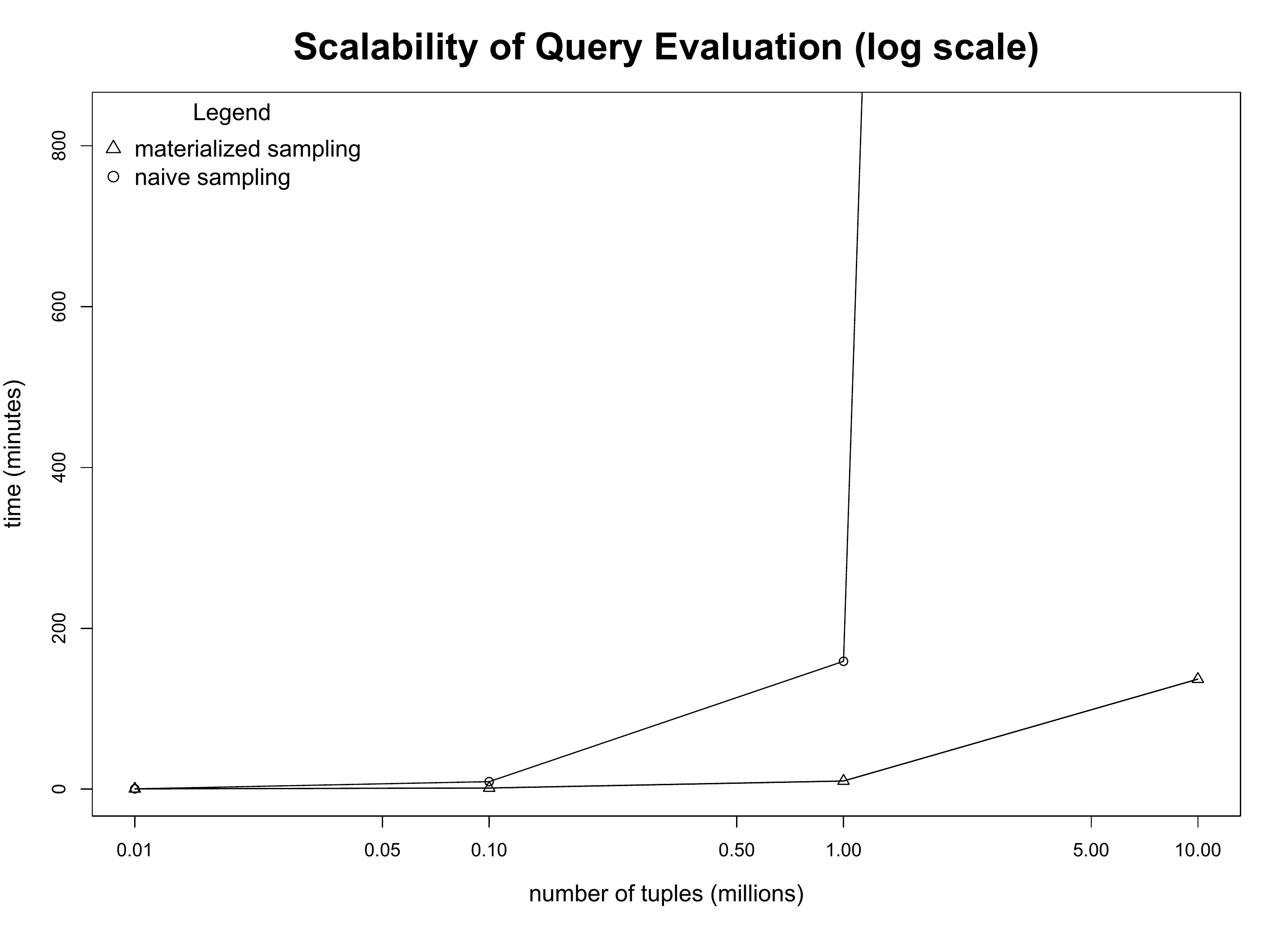} 
}
\hspace{1cm}
\subfigure[Loss versus time comparison for the two query evaluation approaches. Query 1 is evaluated over one-million tuples.] 
{
    \label{fig:hazmat-comparison}
    \includegraphics[width=0.44 \textwidth]{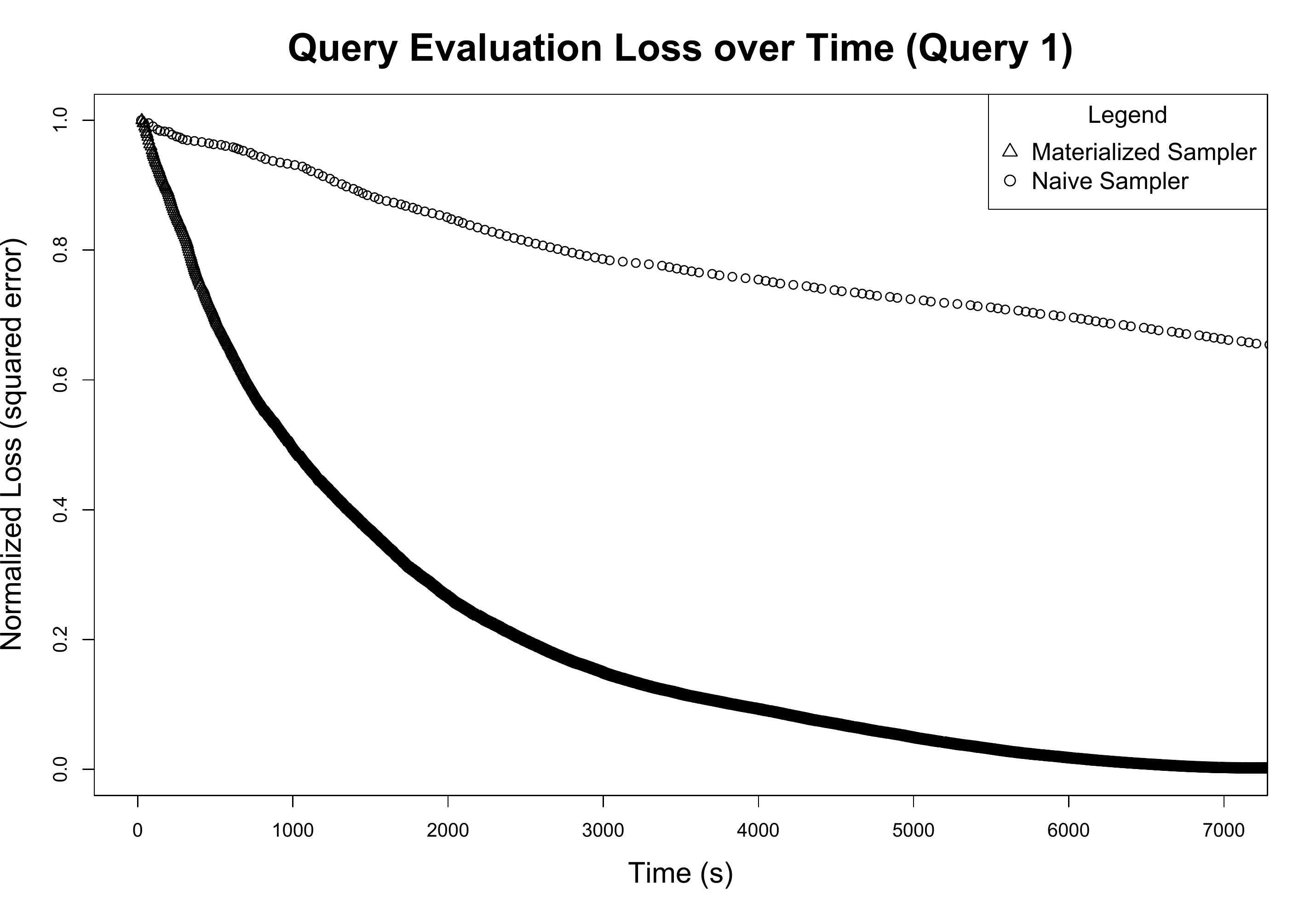}
}
\caption{The benefits view maintenance query evaluation.}
\label{fig:new_materialize} 
\end{figure}

\noindent In Figure~\ref{fig:scalability-query} we plot query
evaluation time versus the number of tuples in the database (log
scale) for both the naive and materialized approach (over several base
ten orders of magnitude). As stated earlier, we have no way of
obtaining the true probabilities, so we estimate the ground truth by
sampling and then define the {\em query evaluation time} as the time
taken to half the quad loss (squared error) from the initial
``single-sample'' deterministic approximation to the query.

For small databases, the sampler based on view-maintenance techniques
does not provide efficiency gains over the naive approach. Indeed,
when the database contains just 10,000 tuples, the two approaches
perform comparably: the naive sampler is slightly quicker (19 seconds
compared to 21 seconds) possibly due to the overhead involved in
maintaining the auxiliary diff tables (recall that the size of the
diff tables is roughly 10,000 tuples because there are that many steps
between query executions). For 100,000 tuples, the view-based approach
begins to outperform the naive approach (162 seconds versus 178 for
naive) and quickly yields dramatic improvements as the number of
tuples increases. In fact, we were unable to obtain the final
data-point (ten million tuples) for the naive sampling approach
because we project it to take 227 hours to complete. In stark
contrast, the sampler based on view-maintenance techniques takes under
two-and-a-half hours on the same ten million tuples. We are impressed
with the speed of the evaluation because inference in skip chains CRFs
is extremely difficult and normally takes hours to complete|even in
the non-database setting.

It is worth noting that for the skip-chain CRF (and the sophisticated
entity-wise coreference model presented in Figure~\ref{fig:pedagog}),
the time to perform an MCMC walk-step is constant with respect to the
size of the database. That is, if the proposal distribution only
modifies a constant number of variables, then only a constant number
of tuples in the database are involved in computing the acceptance
computation (see the Appendix,
Section~\ref{app:efficencies}). Therefore, because the time to perform
a single walk-step is constant with respect to the size of the
repository, only two primary factors affect scalability: (1) the
DBMS's deterministic query execution time and (2) the number of
samples required to change the database in a meaningful way. This
suggests two avenues of future work for improving scalability even
further. In particular investigating jump functions that better
explore the space of possible worlds appears to be an extremely
fruitful venture with high dividends.

Next, in Figure~\ref{fig:hazmat-comparison}, we plot query-evaluation
error versus time for both query evaluators on the 1-million tuple
database. Recall that the two approaches generate the same set of
samples, but the naive approach is slower because it must execute the
query on each possible world (rather than exploiting the set of
modified tuples). Impressively, the efficient evaluator nearly zeroes
the error before the naive approach can even half the error. Also,
notice how loss tends to decrease monotonically over time. This allows
our approach to be used as an any-time algorithm: applications that
require fine probability estimates can spend more evaluation time,
while those that are time sensitive can settle for courser estimates.

\subsection{Parallelization}
In general, sampling approaches to query evaluation can be easily
parallelized to yield impressive performance improvements. These
performance improvements are potentially even greater in the context
of our MCMC approach because parallelization provides the additional
benefit of generating samples with higher independence, leading to
faster mixing rates. In this section we show that running multiple
query evaluators in parallel dramatically improves the accuracy given
a fixed time-span, demonstrating our system's potential to satisfy
high-fidelity requirements in time-sensitive situations.

\begin{figure}
\includegraphics[width=0.49 \textwidth]{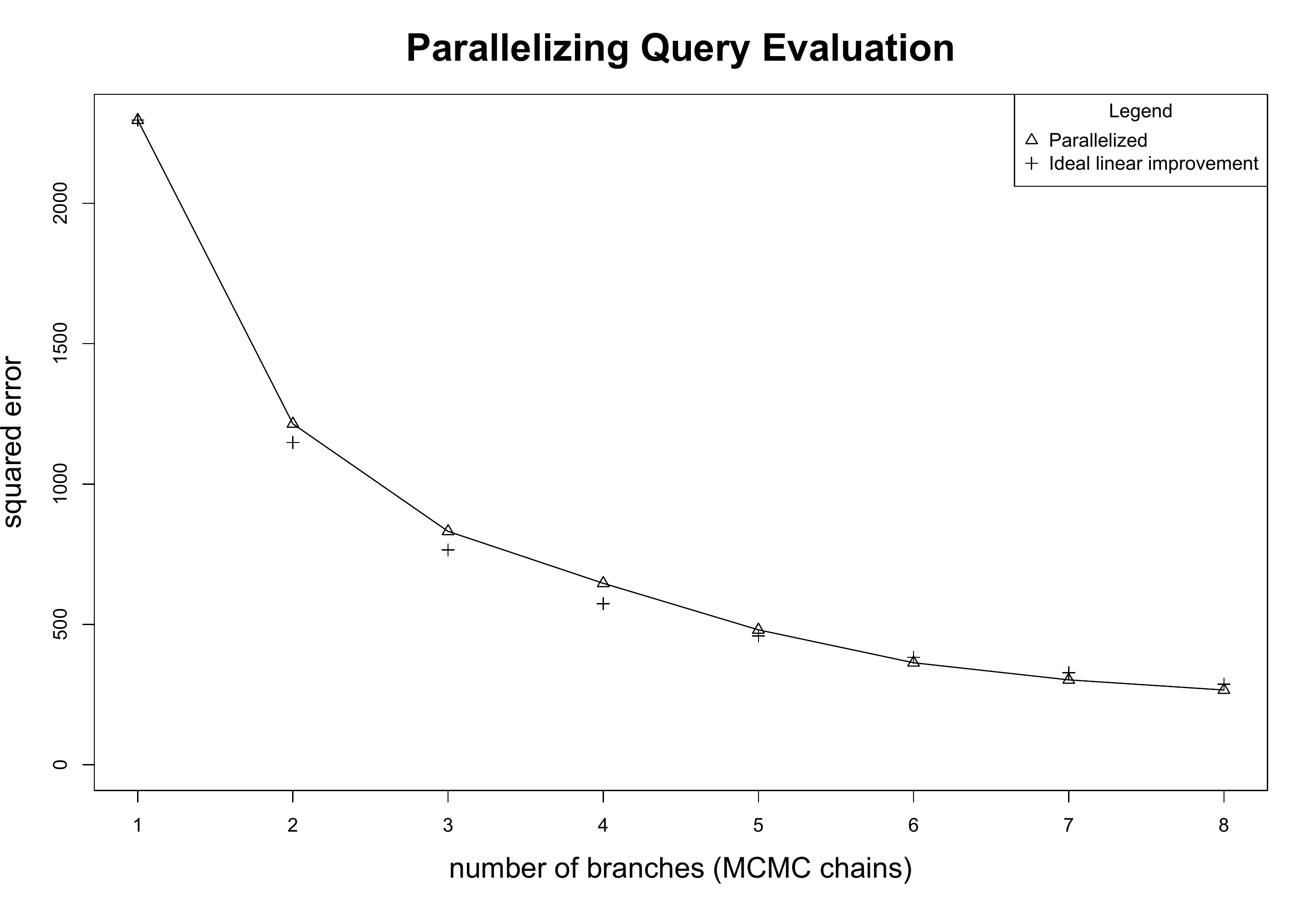}
\caption{Multiple evaluators in parallel.}
\label{fig:parallel}
\end{figure}

We evaluate the effects of parallelization as follows. First we
produce eight identical copies (initial worlds) of the probabilistic
database, each with ten million tuples. We evaluate Query 1 using the
usual set-up except we obtain the ground-truth by averaging eight
parallel chains for ten-thousand samples each. To evaluate the query,
we run up to eight parallel query evaluators for one-hundred samples
(with the usual ten-thousand MCMC steps in between each sample), the
results are plotted in Figure~\ref{fig:parallel} and compared against
the ideal linear improvement. For example, by using two chains we
almost half the loss of the one chain evaluator. Impressively, eight
chains reduces the error by slightly more than a factor of eight,
demonstrating that MCMC sampling evaluation can be further improved by
parallelization.

As we can see, this simple form of parallelization is actually quite
powerful because samples taken across multiple chains are much more
independent than those taken within a single chain. This is one reason
why we actually observe super-linear improvements in fidelity through
parallelization. These benefits come at a relatively small cost of (1)
additional hard-drive space for storing multiple worlds
simultaneously, and (2) additional processors for parallelization.

\subsection{Aggregates}
Another benefit of sampling-based approaches is their ability to
answer arbitrary relational algebra queries without the need to close
a representation system under the necessary operators. In this section
we empirically demonstrate that our system is capable of answering
arbitrary extensions to relational algebra such as aggregates. We
begin with a simple aggregate query that counts the number of person
mentions in one-million tuples worth of New York Times
tokens. \\

\noindent{\bf Query 2} \\
{\small
\indent SELECT COUNT(*) \\
\indent FROM TOKEN \\
\indent WHERE LABEL='B-PER' \\
}

\noindent The second aggregate query retrieves documents in which the
number of {\em person} mentions is equivalent to the number of {\em
  organization} mentions within that document (again applied to
one-million tuples).\\

\noindent{\bf Query 3} \\
{\small
\indent SELECT T.doc\_id\\
\indent FROM Token T \\
\indent WHERE (SELECT COUNT(*) \\
\indent\indent FROM Token T1 \\
\indent\indent WHERE T1.label=`B-PER' AND T.doc\_id=T1.doc\_id) \\
\indent=(SELECT COUNT(*) \\
\indent\indent FROM Token T1 \\
\indent\indent WHERE T1.label=`B-ORG' AND T.doc\_id=T1.doc\_id) \\
} \\

\noindent We plot squared error loss as a function of time for these
two queries. The ground-truth was obtained by running each query for
five-thousand samples with the usual ten-thousand MCMC walksteps
between each sample. We see that Query 2 rapidly converges to zero
loss, and Query 3 converges at a respectable rate. In fact, the rapid
convergence of Query 2 can be explained by examining its answer set,
which we provide in Figure~\ref{fig:skip-agq1-histo} of the
Appendix. Notice how the distribution is highly peaked about the
center and appears to be normally distributed. This is not unusual for
real-world data, and MCMC sampling is celebrated for its ability to
exploit this concentration of measure, leading to rapid convergence.

\begin{figure}[h]
\includegraphics[width=0.49 \textwidth]{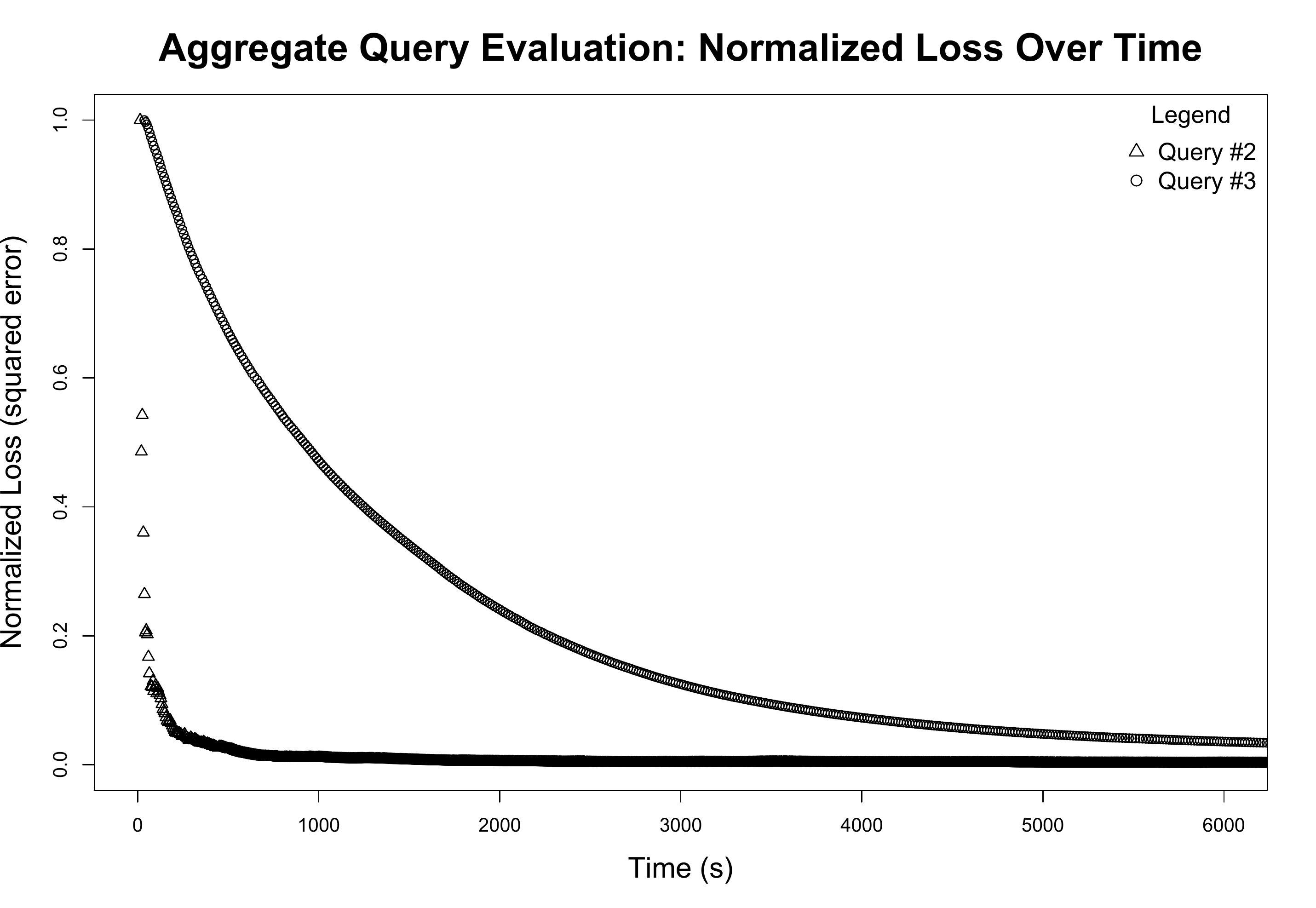}
\caption{Squared loss over time time for two aggregate queries (Query 2 and Query 3), results reported on one-million tuples.}
\label{fig:aggregates}
\end{figure}

\section{Conclusion and Future Work}

In this paper we proposed a framework for probabilistic databases that
uses factor graphs to model distributions over possible worlds. We
further advocated MCMC sampling techniques and demonstrated how the
Markovian nature can be exploited to efficiently evaluate arbitrary
relational queries in an any-time fashion.

\cut{
In this paper we proposed a novel framework for probabilistic
databases based on factor graphs using Metropolis-Hastings sampling to
hypothesize query-able worlds. We implemented a prototype system on a
real-world dataset and demonstrated the efficiency and accuracy of our
methods. We further investigated the importance of the jump function
and observed how bias affects our ability to cache
tuples. Additionally we presented empirical evidence suggesting
query-specific jump functions can lead to large efficiency gains.
}

In future work we would like to investigate methods for {\em
  automatically} constructing jump functions to target specific
queries. \cut{We also wish to use our system to store and jointly model
larger and more relational datasets.}

\section{Acknowledgments}
This work was supported in part by the Center for Intelligent
Information Retrieval, in part by Lockheed Martin prime contract
\#FA8650-06-C-7605 through a subcontract with BBNT Solutions LLC, in
part by SRI International subcontract \#27-001338 and ARFL prime
contract \#FA8750-09-C-0181, in part by The Central Intelligence
Agency, the National Security Agency and National Science Foundation
under NSF grant \#IIS-0326249, and in part by UPenn NSF medium
IIS-0803847. Any opinions, findings and conclusions or recommendations
expressed in this material are the authors' and do not necessarily
reflect those of the sponsor.

\bibliographystyle{abbrv}
\bibliography{pdb-vldb}  

\newpage
\section{Appendix}

\subsection{Examples Probabilistic Query Answers}
Here we provide a few examples of query answers. Recall that answers
contain tuples along with their probabilities. In each of these plots
the $x$ axis ranges over actual tuples and the height of the bar show
the probability of that tuple being in the
answer. Figure~\ref{fig:skip-agq1-histo} shows the answer to Query 2,
an aggregate query asking the number of person mentions (``B-PER''),
over ten million tokens from NYT articles from the year 2004. Notice
that the mass appears to be normally distributed, where the important
observation is that most of the mass is clustered around a small
subset of the answer set. This important property is exhibited by many
real-world datasets, and enables MCMC to rapidly converge to the true
stationary distribution.

\begin{figure}[h]
\includegraphics[width=0.49 \textwidth]{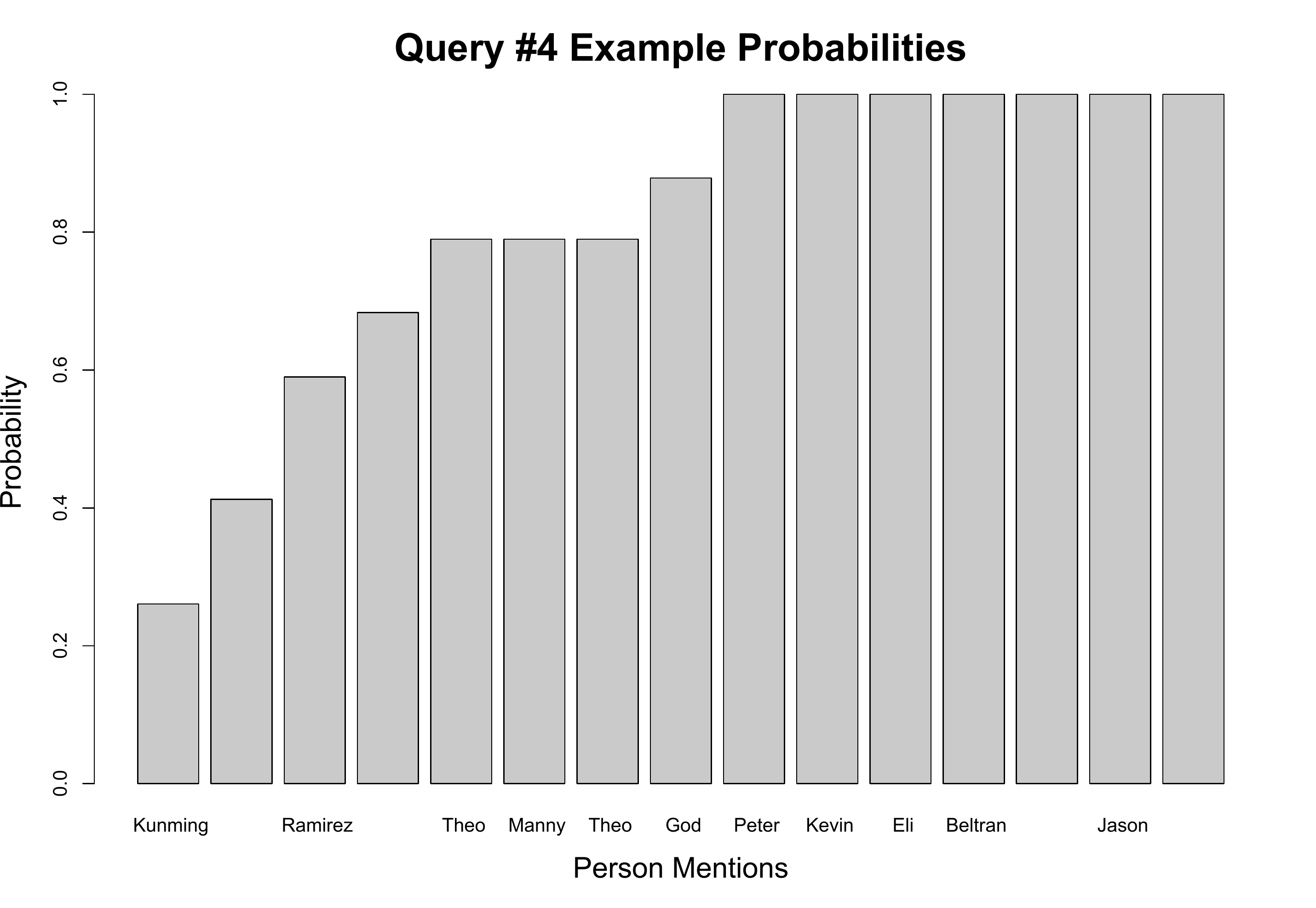}
\caption{Selected tuples from Query 4 over NYT tuples}
\label{fig:skip-boston}
\end{figure}

In Figure~\ref{fig:skip-boston} we show a subset of the answer to
Query 4, which seeks all person mentions (``B-PER'') that co-occur (in
the same document) as a token with string ``Boston'' having label
``B-ORG''. Intuitively, ``Boston'' can ambiguously be a location or an
organization (because organizations are often named after the city in
which they are based). \\

\noindent{\bf Query 4} \\
\indent SELECT T2.STRING \\
\indent FROM TOKEN T1, TOKEN T2 \\
\indent WHERE T1.STRING='Boston' AND T1.LABEL='B-ORG' \\
\indent\indent AND T1.DOC\_ID=T2.DOC\_ID AND T2.LABEL='B-PER'\\


\noindent We find that many of the people returned in our query are
affiliated with baseball likely because the Boston Red Sox are a
prominent example of an organization named after a city.

  \cut{ can
  refer to an organization as in {\em Boston Dynamics} or {\em Boston
    Red Sox}. Indeed we see that many of the people that cooccur with
  instances of Boston that are organizations are names of Red Sox team
  members or other baseball players.}

\begin{figure}[h]
\includegraphics[width=0.49 \textwidth]{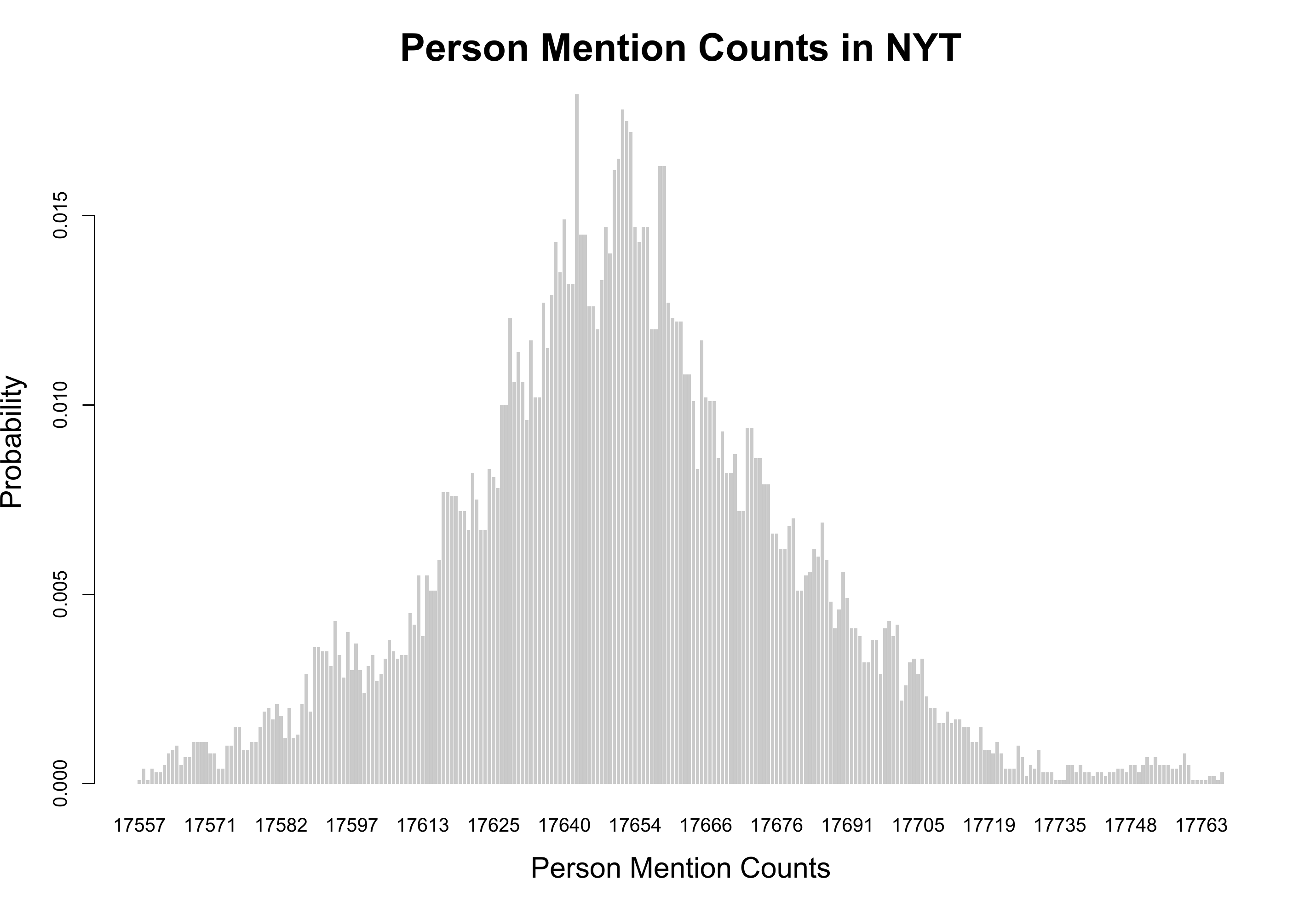}
\caption{Aggregate query (Query 2) distribution as a histogram. Shows
  the distribution of person mention counts over 10 million NYT tuples.}
\label{fig:skip-agq1-histo}
\end{figure}

\subsection{MCMC Efficiencies}
\label{app:efficencies}
The following shows how the acceptance ratio in Metropolis Hastings
can be computed efficiently. We begin with the MH acceptance ratio,
which depends on the probabilities expressed by the factor graph 
(Equation~\ref{eqn:conditional222}). Simple algebraic manipulation
allows the ratio to be expressed in terms factors neighboring only
those variables that change:

\label{sec:app-mh-efficient}
\begin{align*}
  \frac{p(y')}{p(y)} & =\frac{p(Y=y'|X;\theta)}{p(Y=y|X;\theta)} \\
 & =\frac{\frac{1}{Z_X} \prod_{y^i \in y'}{\psi(X,y^i)}}{\frac{1}{Z_X} \prod_{y^ \in y}{\psi(X,y^i)}}\\
 & =\frac{\left(\prod_{y^i \in \delta_{y'}}{\psi(X,y^i)}\right) \left(\prod_{y^i \in y'-\delta_{y'}}{\psi(X,y^i)}\right)}
     {\left(\prod_{y^i \in \delta_y}{\psi(X,y^i)}\right) \left(\prod_{y^i\in y-\delta_y}{\psi(X,y^i)}\right)} \\
 & =\frac{\prod_{y^i \in \delta}{\psi(X,y^i)}}{\prod_{y^i \in \delta}{\psi(X,y^i)}} \\
\end{align*}

For example, take the skip chain conditional random field presented in
Figure~\ref{fig:skip-chain} of Section~\ref{sec:experiments}. Suppose
an initialization where the middle ``IBM'' token is assigned the label
``LOC'' and our jump function proposes to change the label to ``ORG'',
then we only need to compute twelve factors to evaluate the MH
acceptance ratio and decide whether to accept this jump. For this
model and proposal distribution, the number of factors we ever need to
evaluate is constant with respect to the number of tokens in the
database.

\begin{figure}[h]
\includegraphics[width=0.49 \textwidth]{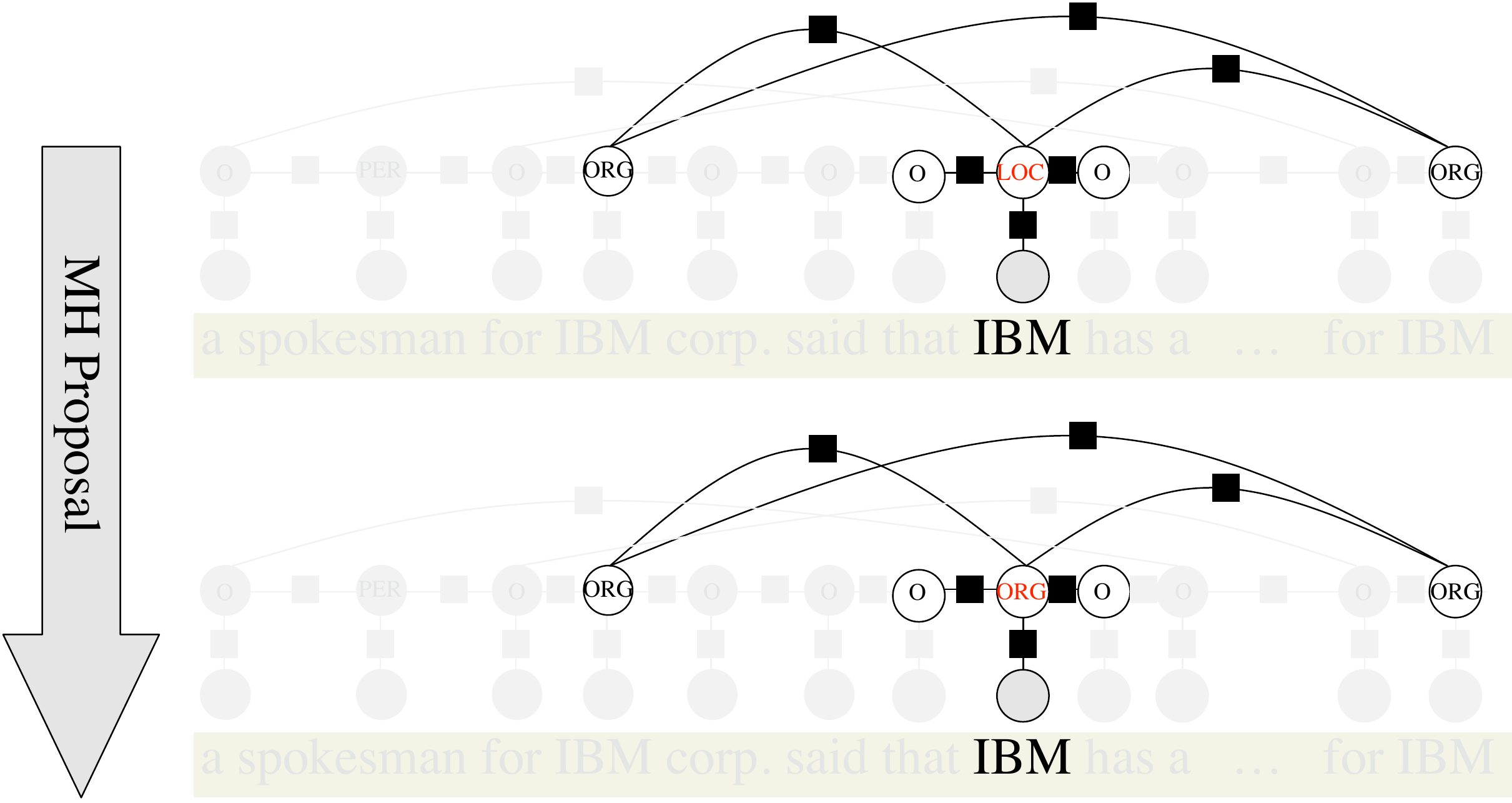}
\caption{Efficient Metropolis-Hastings evaluation for the skip chain
  presented in Section~\ref{sec:experiments}. Greyed out factors
  cancel in the MH proposal ratio and their corresponding greyed-out
  arguments may be ignored. Only black factors need to be evaluated.}
\label{fig:skip-agq1-histo}
\end{figure}

\subsection{BIO Labels for Named Entity Recognition}
\label{sec:app-bio}
BIO labels allow textual mentions to be composed of more than one
token by prefixing the labels with a {\em B-$<$T$>$} indicating that
the token is beginning a mention of type {\em $<$T$>$}, and {\em
  I-$<$T$>$} indicating the token is continuing a mention of type {\em
  $<$T$>$}; and an {\em O} indicating the word is not any type of
mention.

As an example, if we annotate the sentence {\em he saw Hillary Clinton speak} as: \\

\noindent he {\bf (B-PER)}, saw {\bf (O)}, Hillary {\bf (B-PER)}, Clinton {\bf (I-PER)}, speaks {\bf O} \\

\noindent then the sentence is interpreted as having two mentions:
{\em he} and {\em Hillary Clinton}. Note that I-$<$T$>$ can follow
B-$<$U$>$ if and only if $T=U$, otherwise, the interpretation is
meaningless. This suggests we could devise a more intelligent jump
function that takes this constraint into account.

\subsection{Query Evaluation}
Here we show the basic components of query evaluation. First a
Metropolis-Hastings random walk is presented in
Algorithm~\ref{alg:mh}. The algorithm takes an initial world $w_0$,
then executes $n$ proposals, resulting in a random walk, ending in
some final world $w$'.

\begin{algorithm}[t]
  \caption{Random Walk with Metropolis Hastings ($n$ steps)}\label{alg:mh}
  \begin{algorithmic} [1]
    \STATE {\bf Input:} \\Initial world $w$,\\ number of steps $n$
    \FOR{$i=1,\hdots, n$}
       \STATE{$w' \sim q(w)$}
       \IF{$true \sim  \alpha(w',w)$}
          \STATE{$w \leftarrow w'$}
       \ENDIF
       \STATE{{\bf return} $w$}
    \ENDFOR    
  \end{algorithmic}
\end{algorithm}

Next we show the basic query evaluation method
(Algorithm~\ref{alg:basic-sampling}). This method evaluates a query
$Q$ on the probabilistic database. Recall that the database always
stores a single possible world and is initialized to some world
$w$. To collect a sample, $k$ MH walksteps are taken to transition the
database to some new world $w'$. Then the query $Q$ is executed over
this deterministic world and tuple-counts are collected. This process
is repeated $n$ times. Note that this is the basic MH query evaluator
that does not exploit the overlap between consecutive MCMC samples;
the more sophisticated view-maintenance evaluator is described in the
body of this manuscript.

\begin{algorithm}[t]
  \caption{Basic Query Evaluation Method}\label{alg:basic-sampling}
  \begin{algorithmic} [1]
    \STATE {\bf Input:} \\ initial world  $w_0$, \\number of steps $n$ \\ number of samples per query: $k$
    \STATE{{\bf Initialization:}
      \\ //initial state\\$w \leftarrow w_0$
      \\ //initial marginal counts
      \\ ${\bf m}\leftarrow {\bf 0}$
      \\ //initial normalizing constant for marginals
      \\ $z\leftarrow 0$}
    \FOR{$i=1,\hdots, n$}
       \STATE{
         //run MH for $k$ steps beginning on world $w$ \\
         $w\leftarrow $ MetropolisHastings(w,k)}
       \STATE{//run query on sampled world\\
         $s\leftarrow Q(w)$
       }
       
       \STATE{//increase counts \\${\bf m}\leftarrow m_i+
       \begin{cases}
         1 & \text{if $m_i \in s$} \\
         0 & \text{o.w.} \\ 
       \end{cases}$}
       \STATE{$z\leftarrow z + 1$}
    \ENDFOR
    \STATE{{\bf return} $\frac{1}{z} {\bf m}$}
  \end{algorithmic}
\end{algorithm}

\end{document}